\documentclass{WileyMSP-template}
\newcommand\beq{\begin{equation}}
	\newcommand\eeq{\end{equation}}
	
	\usepackage{graphicx}
	\usepackage{amsmath,amssymb}
	\usepackage{amsfonts}
	\usepackage{color}
	\usepackage{cite}
    \usepackage{bm}

\begin{document}



\title{Optical Spin Effects Induced by Phase Conjugation at a Space-Time Interface}
\maketitle


\author{Carlo Rizza}
\author{Alessandra Contestabile}
\author{Maria Antonietta Vincenti}
\author{Giuseppe Castaldi}
\author{Marcello Ferrera}
\author{Alessandro Stroppa}
\author{Michael Scalora}
\author{Vincenzo Galdi*}



\begin{affiliations}
Prof. C. Rizza, A. Contestabile\\
Department of Physical and Chemical Sciences, University of L’Aquila, I-67100 L’Aquila, Italy\\

Prof. G. Castaldi, Prof. V. Galdi\\
Fields \& Waves Lab, Department of Engineering, University of Sannio, I-82100 Benevento, Italy\\
Email Address: vgaldi@unisannio.it\\

Prof. M. A. Vincenti\\
Department of Information Engineering, University of Brescia, I-25123 Brescia, Italy\\

Prof. M. Ferrera\\
Institute of Photonics and Quantum Sciences, Heriot-Watt University, SUPA Edinburgh, \\ EH14 4AS United Kingdom\\

Dr. A. Stroppa\\
CNR-SPIN, c/o Department of Physical and Chemical Sciences, University of L'Aquila, I-67100 L’Aquila, Italy\\

Dr. M. Scalora\\
Aviation and Missile Center, U.S. Army CCDC, Redstone Arsenal, Alabama 35898-5000, USA\\

\end{affiliations}



\begin{abstract}

Electromagnetic temporal boundaries, emerging when the constitutive parameters of a medium undergo abrupt temporal variations, have garnered significant interest for their role in facilitating unconventional wave phenomena and enabling sophisticated field manipulations.  A key manifestation is temporal reflection in an unbounded spatial domain, where a sudden temporal discontinuity induces phase-conjugated backward waves alongside anomalous spin conversion.  This study explores distinctive spin-conversion dynamics at a time-dependent spatial interface governed by Lorentz-type dispersion, in which the plasma frequency undergoes rapid modulation over time.  The interaction of a circularly polarized wave with a space-time interface excites electromagnetic signals at the system’s natural resonance, allowing precise control over polarization states. The scattered field stems from the combined influence of temporal and spatial boundaries, yielding a superposition of the original incident wave’s polarization and its phase-conjugated counterpart.

\end{abstract}

\section{Introduction}
Electromagnetic time-varying media, characterized by material properties that change rapidly over time \cite{Galiffi:2022po}, have attracted significant attention in recent years. While the theoretical framework underpinning this field has been established for decades \cite{Morgenthaler:1958vm,Oliner:1961wp,Felsen:1970wp,Fante:1971to,AuYeung:83}, only in recent years have experimental platforms emerged that enable controlled time-dependent behavior \cite{Kamaraju:2014,Lee:2018,Moussa:2023,Lustig:2023,Jones:2024,Jaffray2025}. By extending classical spatial concepts into the temporal domain, researchers have explored a variety of novel and intriguing configurations. 
\begin{figure}
\begin{center}
\includegraphics[width=0.6\columnwidth]{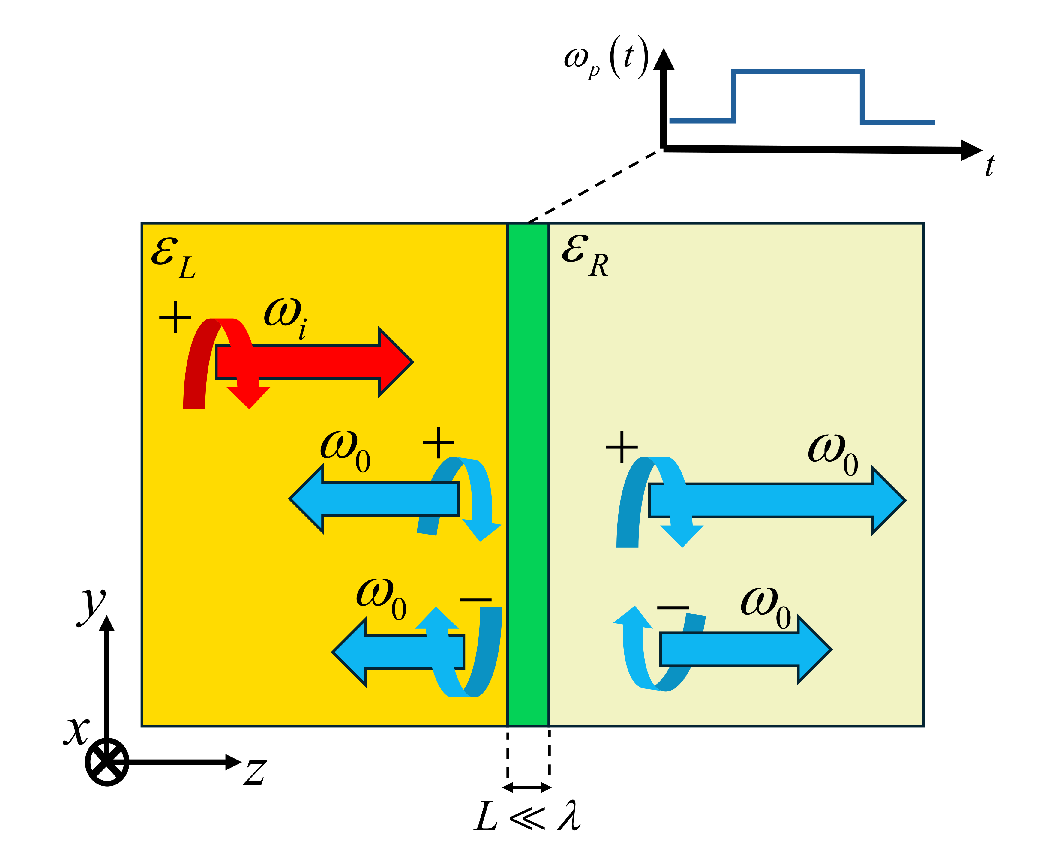}    
\end{center}
\caption{Schematic illustration of the proposed mechanism. A time-varying dispersive interface is embedded between two homogeneous and stationary half-spaces with relative permittivities $\varepsilon_L$ and $\varepsilon_R$. The interface is modeled as a thin slab of thickness $L \ll \lambda$, and exhibits Lorentz-type dispersion with a plasma angular frequency $\omega_p$ that varies in time. When a circularly polarized wave packet with positive spin (i.e., RHCP) and angular frequency $\omega_i$ (red arrow) impinges on the interface, the scattering process generates both LHCP and RHCP components at the resonance angular frequency $\omega_0$ (blue arrows).}
\label{fig1}
\end{figure}
These advancements encompass a broad spectrum of time-varying structures and functionalities, including temporal boundaries \cite{Xiao:2014ra}, time-modulated slabs \cite{Ramaccia:2020}, dynamic filters \cite{Silbiger:2023,Castaldi2022,Galiffi:2022tp}, time-domain antireflection coatings \cite{Pacheco-Pena:20}, photonic time crystals \cite{Lustig:23}, artificial magnetism \cite{Rizza:2022no}, and wave-based analog computing \cite{Rizza:2022sh,Rizza:2023spi}.

By circumventing certain limitations of linear time-invariant systems  \cite{Hayran:2023ut}, such as the constraint of energy conservation and Lorentz reciprocity, time-varying media unlock new avenues for advanced electromagnetic field manipulation. A key phenomenon in these systems is the reflection and refraction at a temporal boundary, which arises when an electromagnetic parameter of a spatially unbounded medium undergoes an abrupt temporal variation. In this context, the scattered waves experience a frequency shift, and the backward-propagating wave exhibits phase conjugation relative to the incident one \cite{AuYeung:83}. Furthermore, the concurrent modulation of both temporal and spatial parameters provides a robust mechanism for achieving comprehensive control over wave propagation \cite{Galiffi:2019br,Galiffi:2022ana,Pacheco-Pena:24,Jajin:2024,Ciabattoni:2025,Bahrami:2025}.

To achieve precise control within practical and compact platforms, a natural approach is to leverage time-varying properties in engineered structures such as metasurfaces. Time-varying metasurfaces have been extensively studied for their ability to facilitate diverse dynamic functionalities, including frequency conversion \cite{Rizza:2024}, optical vortex manipulation \cite{Sedeh:2020}, polarization control \cite{Yang:2021,Hu:2022}, and broadband spectral camouflage \cite{Liu:2019}.

In the context of time-varying media, although the distinctive effect of phase conjugation has recently been experimentally validated \cite{Jones:2024}, the accompanying spin-conversion phenomena remain largely unexplored. Optical phase conjugation is traditionally observed in nonlinear media \cite{Boyd}, where it enables the generation of a time-reversed wavefront. Notably, this process can be harnessed to replicate negative refraction \cite{Pendry:2008,Katko:2010,Aubry:2010,Harutyunyan:2013,Vezzoli:2018,Bruno:2020}. Additionally, an ideal phase-conjugating mirror can effectively mitigate optical aberrations, making it particularly beneficial for laser systems and imaging applications. Unlike conventional mirrors, which invert the handedness of circular polarization, phase-conjugating mirrors reflect circularly polarized beams while preserving their handedness. Temporal boundaries intrinsically support linear phase conjugation, and their characteristics can be further refined through carefully designed space-time architectures. For instance, efficient phase conjugation has been demonstrated in space-time modulated waveguide structures \cite{Yin2022}.

In this work, we investigate a space-time interface that combines both spatial and temporal discontinuities. This type of interface gives rise to distinctive spin-conversion effects when interacting with a circularly polarized wave packet. Specifically, the scattered field can acquire elliptical polarization, resulting from the superposition of components with the same polarization as the incident wave and its phase-conjugated counterpart. The interface under consideration is modeled as a dispersive medium with Lorentz-type characteristics, in which the plasma frequency is rapidly modulated in time. This temporal modulation induces frequency generation near the system’s natural resonance \cite{Rizza:2024}, ultimately leading to an unconventional form of polarization and frequency conversion.
  
Although polarization-dependent effects in time-varying media have previously been explored in systems relying on magnetic bias \cite{Li:2022}, chiral responses \cite{Mostafa:2023}, or anisotropic and bi-anisotropic materials \cite{Yang:2021, Mirmoosa:2024}, our results show that similar phenomena can be realized solely through space-time interfaces, without the need for complex material properties. To the best of our knowledge, this is the first demonstration of polarization control of generated frequencies that does not rely on spatial effects, and the approach is, in principle, applicable across a wide spectral range, from radio to near-infrared.

\section{Results and Discussion}
We consider a time-varying spatial interface consisting of a dielectric layer with deeply subwavelength thickness $L$, positioned between two homogeneous, nondispersive and stationary half-spaces with relative permittivities  $\varepsilon_L$ and $\varepsilon_R$
(see Figure \ref{fig1}). We study the scattering of a wave packet, with electric field given by $\bm{{\mathcal{E}}}(z,t) = {\mathcal{E}}_x(z,t)  \hat{\bf e}_x + {\mathcal{E}}_y(z,t)  \hat{\bf e}_y$, propagating along the $z$-axis and normally incident on the time-varying layer. Here and throughout, $\hat{\bf e}_{\alpha}$ denotes the unit vector in the $\alpha$-direction, with $\alpha = x, y, z$. The interface satisfies the condition $L \ll \lambda$, where $\lambda$ is the characteristic wavelength of the incident radiation. For analytical convenience, we model the interface as an idealized zero-thickness layer located at $z = 0$. Within this framework, the electromagnetic response of the space-time interface is described through a surface polarization $\bm{\mathcal{P}}$, and the resulting electromagnetic interaction is governed by the following equations:
\begin{subequations}
    \begin{eqnarray}
    \Delta \bm{\mathcal{E}}(z,t) &=& 0,\\ 
    {\bf \hat{e}}_z \times \Delta  \bm{\mathcal{H}}(z,t) &=&   \frac{d \bm{\mathcal{P}}(t)}{d t}, 
\end{eqnarray}
	\label{ST}
\end{subequations}
where $\Delta \bm{\mathcal{G}}(z,t) = \bm{\mathcal{G}}(0^+,t) - \bm{\mathcal{G}}(0^-,t)$ with $\bm{\mathcal{G}} = \bm{\mathcal{E}}, \bm{\mathcal{H}}$ \cite{Rizza:2024}. These conditions describe the continuity of the tangential electric field and the discontinuity in the magnetic field due to the presence of a time-varying surface polarization $\bm{\mathcal{P}}(t)$ at the interface.

 Furthermore, we assume that the space-time interface follows a Lorentz-type dispersion, in which the plasma frequency is modulated in time. Under this assumption, the temporal evolution of the surface polarization $\bm{\mathcal{P}}(t)$ is governed by the second-order differential equation
\begin{equation}
	\label{Lor_1}
	\frac{d^2 \bm{\mathcal{P}} }{dt^2}+\gamma \frac{d \bm{\mathcal{P}} }{d t}+\omega_0^2 \bm{\mathcal{P}}=\varepsilon_0  L \omega_p^2(t) \bm{\mathcal{E}}(0,t), 
\end{equation}
where $\omega_0$ is the resonance angular frequency, $\gamma$ is the damping coefficient,  $\omega_p(t)$ denotes the time-dependent plasma angular frequency, and $\varepsilon_0$ is the vacuum permittivity \cite{Rizza:2024}.
 By solving the relevant Maxwell curl equations
\begin{subequations}
\begin{eqnarray}
\nabla \times \bm{\mathcal{E}}&=&-\mu_0 \partial_t \bm{\mathcal{H}},\\
\nabla \times \bm{\mathcal{H}}&=& \partial_t \bm{\mathcal{D}},
\end{eqnarray}
\end{subequations}
where $\mu_0$ is the vacuum  permeability, and the electric displacement field is given by $\bm{\mathcal{D}}=\varepsilon_0 \varepsilon_L \bm{\mathcal{E}}$ for $z<0$, and $\bm{\mathcal{D}}=\varepsilon_0 \varepsilon_R \bm{\mathcal{E}}$ for $z>0$,
 the electric field can be expressed as follows, 
 \begin{equation}
 	\label{E}
 	\bm{\mathcal{E}} (z,t)=\left \{ \begin{array}{ll}
 		\bm{\mathcal{E}}_i\left(t-\displaystyle{\frac{z}{c_L}}\right)+\bm{\mathcal{E}}_r\left(t+\displaystyle{\frac{z}{c_L}}\right), \quad z< 0,\\
 		\bm{\mathcal{E}}_t\left(t-\displaystyle{\frac{z}{c_R}}\right), \quad z> 0,
 	\end{array}
 	\right.
 \end{equation}
 where $\bm{\mathcal{E}}_i$, $\bm{\mathcal{E}}_r$, $\bm{\mathcal{E}}_t$ denote the incident, reflected, and transmitted field, respectively. The quantity $c_{j}=c/\sqrt{\varepsilon_j}$ represents the speed of light in medium $j$, where $j=R,L$, and $c$ is the speed of light in vacuum. 
 
 By combining Equations (\ref{E}) and (\ref{ST}), we obtain:
  \begin{subequations}
  \begin{eqnarray}
 \bm{\mathcal{E}}_t(t)&=& \tau \bm{\mathcal{E}}_i(t)-\frac{Z_0}{\sqrt{\varepsilon_L}+\sqrt{\varepsilon_R}} \frac{d \bm{\mathcal{P}}(t)}{dt},   \\
 \bm{\mathcal{E}}_r(t)&=&(\tau-1) \bm{\mathcal{E}}_i(t) -\frac{Z_0}{\sqrt{\varepsilon_L}+\sqrt{\varepsilon_R}} \frac{d \bm{\mathcal{P}}(t)}{dt}, \quad  
 \end{eqnarray}
 \label{TR}
 \end{subequations}
where $\tau=2 \sqrt{\varepsilon_L}/({\sqrt{\varepsilon_L}+\sqrt{\varepsilon_R}})$ is the transmission (forward-wave) coefficient \cite{Xiao:2014ra}, and $Z_0=\sqrt{\mu_0/\varepsilon_0}$ is the vacuum intrinsic impedance. Moreover, Equation (\ref{Lor_1}) can be rewritten in the form
 \begin{equation}
 	\label{Lor_2}
 	\frac{d^2 \bm{\mathcal{P}}(t) }{dt^2}+\Gamma(t) \frac{d \bm{\mathcal{P}}(t) }{d t}+\omega_0^2 \bm{\mathcal{P}}(t)=\varepsilon_0  \tau L \omega_p^2(t)  \bm{\mathcal{E}}_i(t) , 
 \end{equation}
where we have introduce a time-dependent damping coefficient $\Gamma(t)$, defined as:
 \begin{equation}
 	\Gamma(t)=\gamma+ \frac{L \omega_p^2(t)}{c (\sqrt{\varepsilon_L}+\sqrt{\varepsilon_R})}.
 \end{equation}
In general, Equation (\ref{Lor_2}) does not admit a closed-form analytic solution and must therefore be solved numerically (see Supporting Information for further details). Once the solution for $\bm{\mathcal{P}}(t)$ is obtained, the reflected and transmitted electric fields can be evaluated using Equations (\ref{TR}).
However, in the specific case where a temporal boundary is introduced by an abrupt change in the plasma frequency, Equation (\ref{Lor_2}) can be treated analytically \cite{Rizza:2024}. In this scenario, we assume that the time dependence of the plasma angular frequency is given by
 $\omega_p^2(t)=A(t) \omega_0^2$, 
where 
\begin{equation}
	\label{AA}
	A(t)=A_1  +(A_2-A_1)  U (t),
\end{equation}
and $U(t)$ is the standard unit-step function, defined as $U(t) = 0$ for $t < 0$ and $U(t) = 1$ for $t \geq 0$. The constants $A_1$ and $A_2$ represent the values of the modulation before and after the temporal transition, respectively.
 
In this case, the surface polarization can be expressed as
\begin{equation}
	\label{PPP}
\bm{\mathcal{P}}(t)= U(-t)\bm{\mathcal{P}}_1(t)+ U(t) [\bm{\mathcal{P}}_m (t)+\bm{\mathcal{P}}_2 (t)],
\end{equation}
where $\bm{\mathcal{P}}_j(t)$ represents the steady-state surface polarizations associated with $A(t)=A_j$, for $j=1,2$, and $\bm{\mathcal{P}}_m (t)$ denotes the surface-polarization contribution from the resonance modes supported by the interface for $t>0$.
As will be discussed in detail later, the unexpected term $\bm{\mathcal{P}}_m(t)$ originates from the temporal boundary and plays a central role in enabling the unconventional linear frequency generation, along with the associated spin-conversion effects. To analyze these phenomena, we express each field component using its analytic-signal representation. Specifically, for a generic real-valued field component ${\mathcal{G}}(t)$, we define
\begin{equation}
    {\mathcal{G}}(t)= \textrm{Re} \left[G(t)\right]=\textrm{Re} \left[ \int_{0}^{+\infty}   {\tilde G}(\omega) e^{- i \omega t} d \omega\right],
\end{equation}
where $\tilde{G}(\omega)$ is the one-sided Fourier transform, and $i$ denotes the imaginary unit.

By solving Equation (\ref{Lor_2}) in the steady state for $A(t) = A_j$ (with $j = 1, 2$), we obtain the corresponding spectral representation of the surface polarizations $\bm{\mathcal{P}}_j(t)$, namely, ${\tilde{\bf P}}_j( \omega)= \varepsilon_0 \tau  {\tilde{\chi}}_j(\omega) {\tilde{\bf E}}_i (\omega)$, where 
 \begin{equation}
	\label{P12}
 	{\tilde{\chi}}_j(\omega) =   \frac{ \omega_0^2 L }{\omega_0^2-\omega^2-i \omega \Gamma_j} A_j
 \end{equation}
 is the frequency-domain susceptibility associated with the $j$-th steady-state configuration, and
\begin{equation}
\Gamma_j=\gamma+ \frac{\omega_0^2 L}{c (\sqrt{\varepsilon_L}+\sqrt{\varepsilon_R})} A_j.
\end{equation}
The surface polarization associated with the resonance modes corresponds to the solution of the homogeneous form of Equation (\ref{Lor_2}), obtained by removing the non-homogeneous term proportional to the incident electric field $\bm{\mathcal{E}}_i(t)$, and assuming $\omega_p(t) = \omega_0 A_2$. As a result, the resonance-induced surface polarization is given by $\bm{\mathcal{P}}_m(t) = \mathrm{Re}\left[ \mathbf{P}_m(t) \right]$, where
\begin{equation}
	\label{Pm}
	{\bf P}_m(t)=( {\bf P}_I 
+{\bf P}_{II}^*)e^{-i\Omega_0 t-\frac{\Gamma_2}{2}t } ,
\end{equation}
with $\Omega_0= \sqrt{\omega_0^2-\Gamma_2/2}$, and ${\bf P}_{I}$, ${\bf P}_{II}$ denoting constant vectors. 
In accordance with Equation (\ref{Lor_2}), we enforce continuity conditions on the surface polarization and its derivative at the temporal boundary $t=0$ \cite{Solis:2021}, i.e.,
\begin{subequations}
\begin{eqnarray}
			\bm{\mathcal{P}}(0^+)&=&\bm{\mathcal{P}}(0^-),\\
			\frac{d \bm{\mathcal{P}}(0^+)}{dt}&=&\frac{d \bm{\mathcal{P}}(0^-)}{dt}.
\end{eqnarray}
\label{bou}
\end{subequations}
As a result, we obtain the expressions for ${\bf P}_{I}$, and ${\bf P}_{II}$ as a function of the incident electric field. These are given by:
\begin{subequations}
\begin{eqnarray}
	{\bf P}_I &=& \frac{\varepsilon_0 \tau}{2 \Omega_0} \int_{0}^{+\infty}  \left(\Omega_0+\omega+i \frac{\Gamma_2}{2}   \right) \Delta \tilde{\chi}(\omega) {\tilde {\bf E}}_i(\omega) d \omega, \\
	{\bf P}_{II}& =& \frac{\varepsilon_0 \tau}{2 \Omega_0} \int_{0}^{+\infty} \left(\Omega_0-\omega-i \frac{\Gamma_2}{2}   \right) \Delta\tilde{\chi}(\omega) {\tilde {\bf E}}_i(\omega) d \omega,
\end{eqnarray}
\end{subequations}
with $\Delta \tilde{\chi}=\tilde{\chi}_1-\tilde{\chi}_2$.
To gain physical insight into the unusual spin-conversion effects, we examine the resonance eigenwaves associated with the surface polarization $\bm{\mathcal{P}}_m(t)$ in the low-loss limit, where $\Gamma_2 \rightarrow 0$. Furthermore, we assume that the incident wave packet possesses a well-defined polarization state, described by the complex Jones vector $\mathbf{E}_i(t) = E_i(t) \hat{\mathbf{J}}_i$, where
\begin{eqnarray}
	\label{Ei}
	{\bf \hat{J}}_i=\cos{\varphi_i} \hat{\bf e}_x+\sin{\varphi_i} e^{i \delta_i} \hat{\bf e}_y.
\end{eqnarray}
Here, $\varphi_i$ and $\delta_i$ represent the polarization rotation and phase angles, respectively. In this case, the electric field  $ {\bf E}_m(t)$ associated with ${\bf P}_m(t)$  takes the form of a monochromatic signal, given by:
\begin{eqnarray}
	\label{Em}
	{\bf E}_m(t) \simeq (E_I {\bf \hat{J}}_i +E_{II}^*  {\bf \hat{J}}_i^*) e^{-i \omega_0 t},
\end{eqnarray}
where $E_I$, $E_{II}$ are constant amplitudes coefficients, whose values depend on the spectral profile of the incident wave packet. Specifically, they are given by:
\begin{subequations}
\begin{eqnarray}
	E_I &=& \frac{i \tau }{ 2 c (\sqrt{\varepsilon_R}+\sqrt{\varepsilon_L})}  \int_0^{+\infty} \Delta \tilde{\chi} (\omega_0+\omega) \tilde{E}_i(\omega)d\omega, \\
	E_{II} &=& \frac{ i \tau }{ 2 c (\sqrt{\varepsilon_R}+\sqrt{\varepsilon_L})}  \int_0^{+\infty} \Delta \tilde{\chi} (\omega_0-\omega) \tilde{E}_i(\omega)d\omega. 
\end{eqnarray}
	\label{E_I_II}
\end{subequations}

The eigenwave at the resonance angular frequency $\omega_0$, as described by Equation (\ref{Em}), generally exhibits elliptical polarization. The rotation and phase angles of this polarization are determined by the polarization state of the incident wave packet. As shown in Equation (\ref{E_I_II}), the complex amplitudes $E_I$ and $E_{II}$
depend both on the spectral profile of the incident field $\tilde{E}_i(\omega)$ and on the temporal boundary, through 
the difference in susceptibilities $\tilde{\chi}_1- \tilde{\chi}_2$. In the specific case where the spectral content of the incident wave packet is negligible around $\omega_0$, the electric field given in Equation (\ref{Em}) represents the transmitted and reflected components generated at the resonance angular frequency $\omega_0$. Thus, the excited eigenwaves can propagate both forward and backward, allowing spin conversion to be observed in both reflection and transmission.

Therefore, Equation (\ref{Em}) represents the central result of this study, clearly demonstrating both the generation of a signal at the resonance angular frequency $\omega_0$ and the emergence of the phase-conjugated component induced by the temporal boundary. It is important to emphasize that the presence of the phase-conjugated term, as shown in Equation (\ref{Pm}), is a fundamental characteristic of a temporal boundary. In the conventional scenario where an electromagnetic wave packet propagates through a spatially unbounded medium, a temporal discontinuity gives rise to a negative-frequency component (corresponding to a phase-conjugated backward-propagating wave) since the wavevector remains conserved across the temporal interface. 

For example, when a circularly polarized wave encounters a temporal boundary, the resulting backward-propagating wave retains the same polarization handedness as the incident one. This behavior contrasts with conventional spatial reflection, where the reflected wave exhibits the opposite handedness. In the case of our time-varying interface, the incident signal interacts with both spatial and temporal boundaries, producing reflected and transmitted fields that are superpositions of the original polarization state and its phase-conjugated counterpart.

By considering circularly polarized incident waves, where ${\bf \hat{J}}_i=\hat{\bf e}_{\pm}$ with $\hat{\bf e}_{\pm}=(\hat{\bf e}_x \pm i\hat{\bf e}_y)/\sqrt{2}$), we observe that ${\bf \hat{J}}_i=\hat{\bf e}_{\pm}$ and its complex conjugate ${\bf \hat{J}}_i^*=\hat{\bf e}_{\mp}$ form a pair of orthogonal Jones vectors. These vectors constitute a natural basis for the space of all polarization states. This observation leads to the conclusion that, besides its role as a frequency converter, a space-time interface can also function as a compact and ultrafast polarization converter.   

To provide insight into the polarization-conversion effects previously discussed, we present a series of representative examples derived from our analytic model. In the following, we focus our attention on the reflection spectra and we adopt a notational convention in which ``positive'' and ``negative'' spin (denoted by $+$ and $-$, respectively) refer to circular polarization states defined by the unit vectors $\hat{\mathbf{e}}_+$ and $\hat{\mathbf{e}}_-$, respectively. Depending on the direction of wave propagation, these states correspond to either rigth-handed or left-handed circular polarization (RHCP or LHCP, respectively).

As a first case, we consider an incident signal consisting of a RHCP Gaussian wave packet, with its electric field described by the following profile:
\begin{equation}
\label{inc}
{\bf E}_i(t)=E_0 e^{-(t-t_d)^2/\sigma^2 - i\omega_i (t-t_d)} \hat{\bf e}_+, 
\end{equation}
where $E_0$ is a constant amplitude, $\omega_i$ is the carrier angular frequency, $\sigma$ represents the temporal width, and $t_d$ denotes the time delay relative to the temporal boundary at $t=0$.  

\begin{figure*}
 \begin{center}
 	\includegraphics[width=1\textwidth]{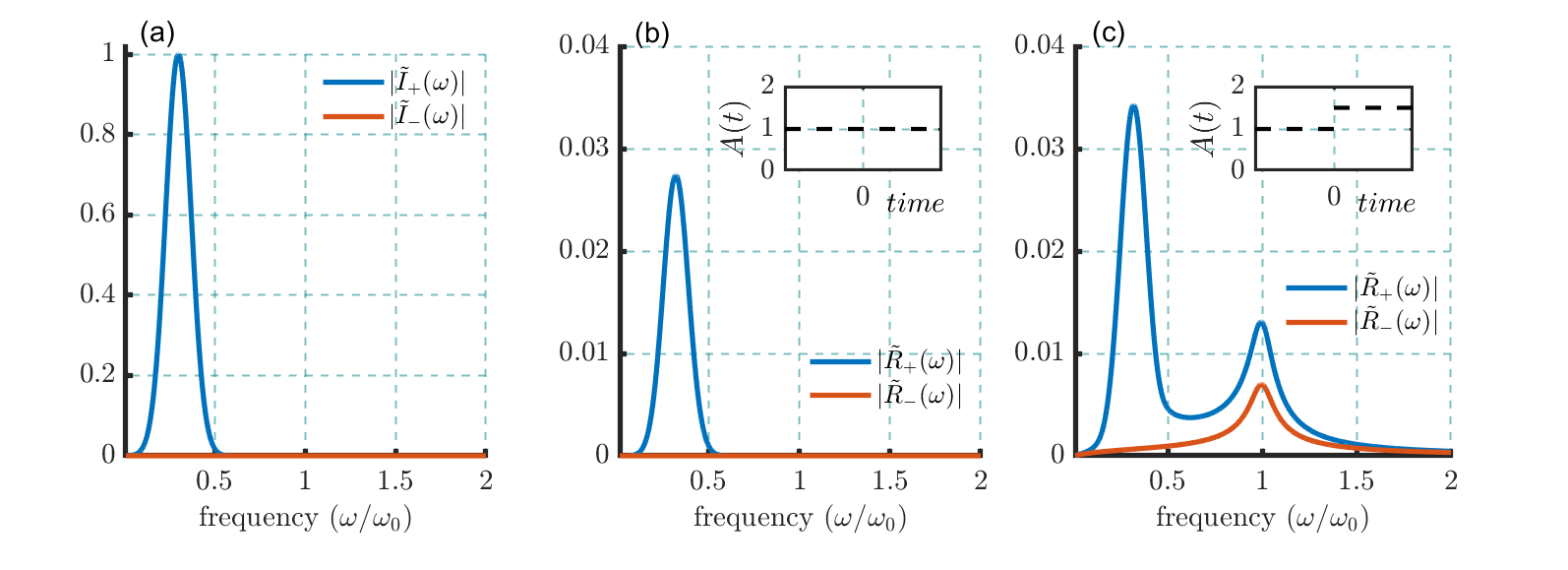}
\end{center}
    \caption{
Example of electromagnetic scattering, for $\varepsilon_L=\varepsilon_R=1$, $L=0.16 c/\omega_0$, $\gamma=5 \cdot 10^{-3} \omega_0$, $\omega_i=0.3 \omega_0$, $\sigma=20/\omega_0$, $t_d=0$, and $A_1=1$. a) Positive- $(\tilde{I}_+$) and negative-spin ($\tilde{I}_-$) spectra of incident wave. b,c) Corresponding spectra ($\tilde{R}_+$ and $\tilde{R}_-$) of  reflected wave in the absence ($A_2 = A_1$) and presence of a temporal boundary ($A_2 = 1.5 A_1$), respectively, as indicated in the insets. All spectra are normalized to the peak value of the incident one.}
 	\label{fig2}
 \end{figure*}

Figure \ref{fig2} illustrates spin-conversion effects by analyzing the reflected wave, using the following parameters: $\omega_i=0.3 \omega_0$, $\sigma=20/\omega_0$, $t_d=0$, $\varepsilon_L=\varepsilon_R=1$, $L=0.16 c/\omega_0$, and $\gamma=5 \cdot 10^{-3} \omega_0$. 
Figure \ref{fig2}a displays the positive- and negative-spin spectra of the incident field, defined as $\tilde{I}_{\pm}(\omega)={\tilde{\bf E}}_i(\omega)  \cdot \hat{\bf e}_{\pm}^*$. Figure \ref{fig2}b and \ref{fig2}c show the corresponding spectral components of the reflected-wave, $\tilde{R}_{\pm}(\omega)={\tilde{\bf E}}_r(\omega) \cdot \hat{\bf e}_{\pm}^*$, in the absence ($A_2=A_1=1$) and presence ($A_2=1.5, A_1=1$) of a  temporal boundary, respectively. In the standard, time-invariant case shown in Figure \ref{fig2}b, the spin state is preserved, and the reflected signal retains a positive spin (LHCP), consistent with the incident polarization. In contrast, Figure \ref{fig2}c shows that when the plasma frequency abruptly changes, a significant spectral component emerges  at the resonance angular frequency $\omega_0$, and remarkably,  components with opposite spin are generated, highlighting the spin-conversion effect induced by the temporal boundary.    

In the Supporting Information, 
the results presented above are validated through full-wave numerical simulations, showing a very good agreement (see Figure S1). Additionally, we examine a scenario where the plasma frequency undergoes a smooth transition following a hyperbolic-tangent profile, characterized by a rising time $t_A$. We find that when $t_A$ is small but finite (specifically, $t_A \lesssim \sigma / 30$), the results remain in close agreement with those obtained assuming an ideal, abrupt temporal boundary (see Figure S2).

From Figure \ref{fig2}, it is evident that the scattered field exhibits a frequency-dependent elliptical polarization, which can be characterized using the frequency-dependent polarization rotation angle $\varphi_j(\omega)$ and phase angle $\delta_j(\omega)$ angle, defined as follows: 
\begin{eqnarray}
	\label{Ert}
	{ \tilde{\bf E}}_j (\omega)= \tilde{E}_j (\omega) 
	\left[\cos{\varphi_j}(\omega) \hat{\bf e}_x+\sin{\varphi_j}(\omega) e^{i \delta_j(\omega)} \hat{\bf e}_y\right],
\end{eqnarray}
where $j=r$, and $j=t$ denote the reflected and transmitted wave, respectively. 

\begin{figure}
\begin{center}
    \includegraphics[width=0.5\textwidth]{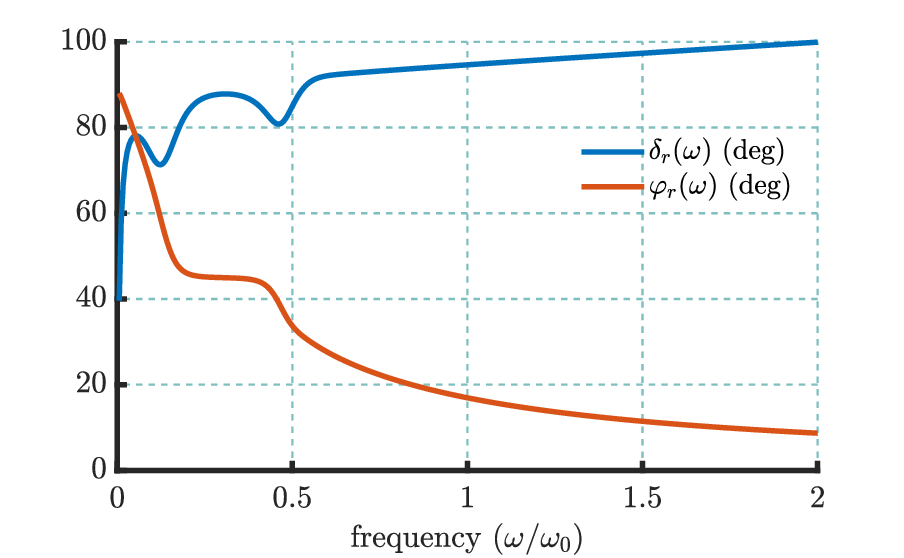}
\end{center}
	\caption{Frequency-dependent polarization rotation angle $\varphi_r$ and phase angle $\delta_r$ of  reflected wave, corresponding to the spectral components shown in Figure \ref{fig2}.}
	\label{fig3}
\end{figure}

Figure \ref{fig3} shows the polarization rotation angle $\varphi_r(\omega)$ and phase angle $\delta_r(\omega)$ of the reflected (backward) wave a function of frequency, using the same configuration as in Figure \ref{fig2}. The results clearly indicate that the reflected field exhibits an unusual, frequency-dependent polarization state. Notably, at the resonance angular frequency $\omega_0$, the phase angle reaches approximately $\delta_r \approx 95^\circ$, while the rotation angle is $\varphi_r \approx 17^\circ$.

\begin{figure*}
	\includegraphics[width=1\textwidth]{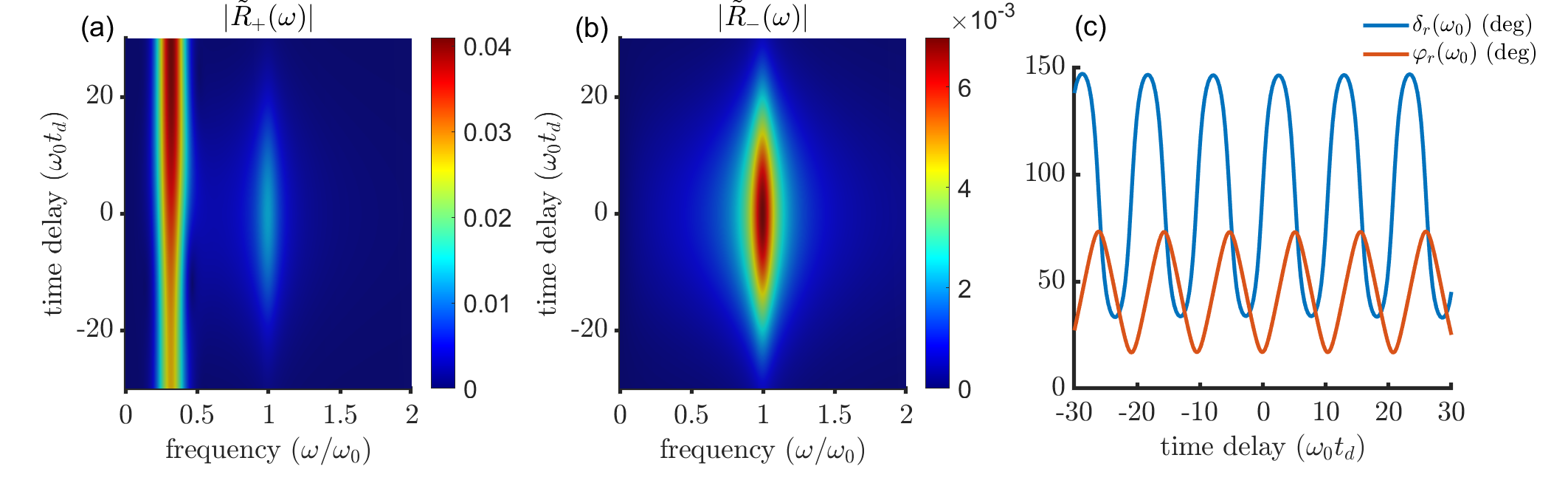}
	\caption{
a,b) Positive- ($\tilde{R}_{+}$) and negative-spin  ($\tilde{R}_{-}$) spectra, respectively, of reflected wave as a function of  time-delay $t_d$.
c) Polarization rotation angle $\varphi_r$ and phase angle $\delta_r$ of reflected (backward) wave at  resonance angular frequency $\omega_0$, as a function of $t_d$.
Incident field and all parameters are identical to those in Figure \ref{fig2}, except for the variation in $t_d$. All spectra are normalized to the peak value of the incident one.}
	\label{fig4}
\end{figure*}
As expected, the observed effect strongly depends on the synchronization between the incident wave packet and the temporal boundary. Figures \ref{fig4}a and \ref{fig4}b display the positive- and negative-spin spectra, respectively, as a function of the time delay $t_d$, using the same parameters as in Figure \ref{fig2}. When $t_d = 0$, the temporal boundary overlaps with the peak of the wave packet, resulting in efficient frequency generation and maximal polarization conversion. In contrast, when $|t_d|$ becomes large, the incident wave arrives either too early or too late relative to the temporal discontinuity, and the effect of the boundary becomes negligible, as clearly illustrated in Figures \ref{fig4}a and \ref{fig4}b. This demonstrates that the time delay $t_d$ serves as a tunable optical \textit{knob} for controlling the polarization state.

Figure \ref{fig4}c shows the polarization rotation and phase angles at the resonance angular frequency $\omega_0$ as a function of $t_d$, using the same parameters as in panels (a) and (b). The results confirm that both angles can be tuned over a relatively wide range, enabling dynamic control of polarization characteristics.

\begin{figure*}
	\includegraphics[width=1\textwidth]{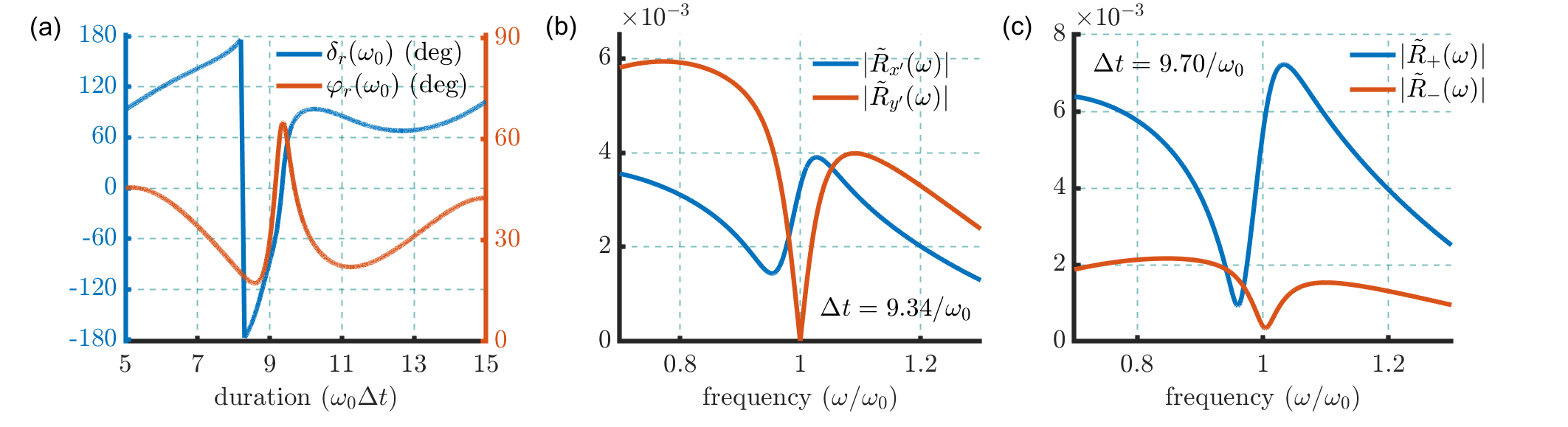}
	\caption{
    Scattering of a positive-spin-polarized field from a space-time interface featuring a temporal slab, defined by $A(t)$ as given in Equation (\ref{AAA}), with $\varepsilon_L=\varepsilon_R=1$, $L=0.16 c/\omega_0$, $\gamma=5 \cdot 10^{-3} \omega_0$, $\omega_i=0.3 \omega_0$, $\sigma=20/\omega_0$, and $t_d=0$.
a) Polarization rotation angle $\varphi_r$ and phase angle $\delta_r$ of reflected wave at the resonance angular frequency $\omega_0$, as a function of the temporal slab width $\Delta t$.
b) Linear-polarization  spectra of reflected wave in the rotated reference system ($\tilde{R}_{x^\prime}$ and $\tilde{R}_{y^\prime}$; details in the text), for a temporal slab with $\Delta t = 9.34/\omega_0$. c)  Circularly polarized spectra of reflected wave for a temporal slab with $\Delta t = 9.70/\omega_0$. All spectra are normalized to the peak value of the incident one.}
	\label{fig5}
\end{figure*}

As a second class of examples, we examine the interaction of the incident field defined in Equation (\ref{inc}) with a time-varying spatial interface that supports a temporal slab, meaning the plasma frequency varies in time according to a soft rectangular function. Specifically, we consider the following configuration:
\begin{equation}
	\label{AAA}
A(t)=1+ \Delta A \left[\tanh{\left(\frac{t}{t_A}\right)} -  \tanh{\left(\frac{t-\Delta t}{t_A}\right)}\right],
\end{equation}
where $\Delta A=0.25$, $\Delta t$ denotes the temporal width of the rectangular pulse, and the rise/fall time is chosen as $t_A=\sigma/20$.

In this more complex scenario, we resort to a semi-analytic approach based on the numerical solution of Equation (\ref{Lor_2}) (see Supporting Information for details). Analogous to a conventional spatial slab, a temporal slab embedded in a time-varying, spatially unbounded medium supports Fabry–Pérot-like resonances and can function as a temporal anti-reflection coating \cite{Ramaccia:2020}. By combining the effects of the spatial interface with those of the temporal slab, it is possible to achieve enhanced control over the polarization state of the scattered field.

Figure \ref{fig5}a illustrates the polarization rotation and phase angles of the reflected wave at the resonance angular frequency $\omega_0$ as a function of the temporal width $\Delta t$. The phase angle $\delta(\omega_0)$ spans the full range from $-180^\circ$ to $180^\circ$, while the rotation angle $\varphi(\omega_0)$ varies from $0^\circ$ to $65^\circ$, indicating strong tunability. In particular, we note that the phase angle $\delta(\omega_0)$ is $\simeq 180^\circ$ and $\simeq 0^\circ$ at $\Delta t = 8.30/\omega_0$, and $\Delta t = 9.34/\omega_0$, respectively, thereby implying that the reflected wave is linearly polarized at the resonance frequency $\omega_0$. Figure \ref{fig5}b displays the linear polarization components of the reflection spectrum [$\tilde{R}_{x'} (\omega)$ and $\tilde{R}_{y'} (\omega)$] for  $\Delta t = 9.34/\omega_0$.
The reflection coefficients $\tilde{R}_{x'} (\omega)$ and $\tilde{R}_{y'} (\omega)$ are expressed in the basis $\hat{\bf e}_{x'}=\cos \phi \hat{\bf e}_x+\sin \phi \hat{\bf e}_y$, $\hat{\bf e}_{y'}=-\sin \phi \hat{\bf e}_x+\cos \phi \hat{\bf e}_y$, where $\phi=\varphi_r(\omega_0) \simeq 64.5^\circ$. As shown in Figure \ref{fig5}b, the $y'$-component of the reflected field is suppressed at $\omega_0$, resulting in a reflected wave that is linearly polarized along the $x'$-axis. Also these results are validated against full-wave numerical simulations in the Supporting Information (see Figure S3). 
Furthermore, with appropriate optimization, it is possible to achieve an almost perfectly circularly polarized state. As an example, Figure \ref{fig5}c shows the circular polarization spectrum of the reflected wave for $\Delta t = 9.70/\omega_0$, clearly demonstrating that the negative spin-state contribution is strongly suppressed at the resonance frequency.

\section{Conclusion}
In conclusion, we have explored unconventional optical spin effects at a space-time interface exhibiting Lorentz-type dispersion, where the plasma frequency is rapidly modulated in time. We have demonstrated that a circularly polarized wave interacting with such an interface can excite waves at the system’s natural resonance, allowing dynamic control over polarization states.

Remarkably, the observed polarization conversion is achieved without invoking bi-anisotropy, chirality, or nonlinear effects.

By focusing on a deeply subwavelength space-time interface, we have developed an analytic framework to characterize the resulting electromagnetic scattering. Importantly, the mechanisms responsible for polarization conversion are not confined to the thin-interface approximation. Instead, the effect leverages the phase conjugation induced by the temporal boundary. Thus, the simultaneous interaction of the electromagnetic field with both temporal and spatial boundaries leads to the generation of scattered waves with elliptical polarization. This occurs because the scattered field is a coherent superposition of waves with incident circular polarization and their phase-conjugated counterpart with opposite handedness.
These results suggest that similar spin-dependent scattering phenomena could be realized in a broader class of time-varying systems, potentially with even greater efficiency. Ultimately, our findings open new avenues for the design of advanced photonic platforms capable of ultrafast control over spin and angular momentum.

From a practical perspective, 
the parameters considered in our study are well within the capabilities of current semiconductor technologies, including GaAs- and Si-based platforms operating at terahertz frequencies. In such materials, femtosecond optical pulses can induce ultrafast changes in the plasma frequency on timescales of a few hundred femtoseconds \cite{Lee:2018,Kamaraju:2014,Shi:2008}. Additionally, precise synchronization between an incident THz wave packet and the induced time-modulated boundary is achievable under realistic experimental conditions \cite{Lee:2018}. Within an optical-pump–THz-probe configuration, a polarized THz wave can undergo phase conjugation at the space–time interface, thereby enabling the observation of the optical spin effects predicted in this work. These considerations indicate that experimental implementation is feasible and will be pursued in future investigations.



\medskip
\textbf{Supporting Information} \par 
Supporting Information is available from the Wiley Online Library.

\medskip
\textbf{Acknowledgements} \par 
The research was partially supported by the European Union -- NextGenerationEU under the Italian MUR National Innovation Ecosystem grant ECS00000041 - VITALITY - CUP E13C22001060006.
The work of G. C. and V. G. was partially supported by the European Union-Next Generation EU under the Italian National Recovery and Resilience Plan (NRRP), Mission 4, Component 2, Investment 1.3, CUP E63C22002040007, partnership on ``Telecommunications of the Future'' under Grant PE00000001 - program ``RESTART''. 

\medskip

%
\bibliographystyle{MSP_titles}
\bibliography{biblio}





\end{document}




\title{Supporting Information\\~\\Optical Spin Effects Induced by Phase Conjugation at a Space-Time Interface}

\maketitle


\author{Carlo Rizza}
\author{Alessandra Contestabile}
\author{Maria Antonietta Vincenti}
\author{Giuseppe Castaldi}
\author{Marcello Ferrera}
\author{Alessandro Stroppa}
\author{Michael Scalora}
\author{Vincenzo Galdi*}





\vspace{2cm}
This document provides some details on numerical simulations, as well as some additional results. Newly introduced equations are labeled with the prefix ``S''; all others pertain to the main text. All bibliographic references are intended as local, and those already utilized in the main text are repeated.

\section{Details on full-wave simulations}
To independently validate our analytic findings, we perform full-wave numerical simulations using the finite element-based commercial software COMSOL Multiphysics. The simulations are conducted in a one-dimensional spatial domain, corresponding to the setup illustrated in Figure \ref{MT-fig1}. Specifically, we consider an incident left-handed circularly polarized (LHCP) wave (positive spin) impinging on a spatial slab of finite thickness $L$. As a result, the computational domain consists of a time-varying slab positioned between two vacuum regions.

In the vacuum regions, the electric field evolves according to the standard wave equation. Within the time-varying slab, however, the field dynamics is governed by a modified wave equation that accounts for the coupling between the electromagnetic wave and the time-varying material properties. This modified equation is expressed as:
\begin{equation}
\label{mod_DA}
\frac{1}{c^2} \frac{\partial^2   \bm{\mathcal{E}}(z,t)}{\partial t^2}-\frac{\partial^2   \bm{\mathcal{E}}(z,t)}{\partial z^2}=- \mu_0 \frac{\partial^2   \bm{\mathcal{P}}_v(z,t)}{\partial t^2}, 
\end{equation}

where $\bm{\mathcal{P}}_v(z,t)$ denotes the electric polarization vector within the slab, whose temporal dynamics are governed by the following differential equation:
\begin{equation}
\label{Lor_2v}
   \frac{d^2 \bm{\mathcal{P}}_v}{dt^2}+\gamma \frac{d \bm{\mathcal{P}}_v}{d t}+\omega_0^2 \bm{\mathcal{P}}_v=\varepsilon_0  \omega_p^2(t) \bm{\mathcal{E}}. 
\end{equation}

We solve Equation (\ref{mod_DA}) in conjunction with Equation (\ref{Lor_2v}) using the PDE module in COMSOL Multiphysics. The simulations are carried out with the standard transient solver, employing default settings and a relative tolerance set to $10^{-6}$. Standard scattering boundary conditions are applied at both the input and output interfaces of the computational domain. For spatial discretization, the mesh is configured with a minimum element size of $0.1L$, ensuring accurate resolution across the slab region.

 \begin{figure*}
 	\includegraphics[width=1\textwidth]{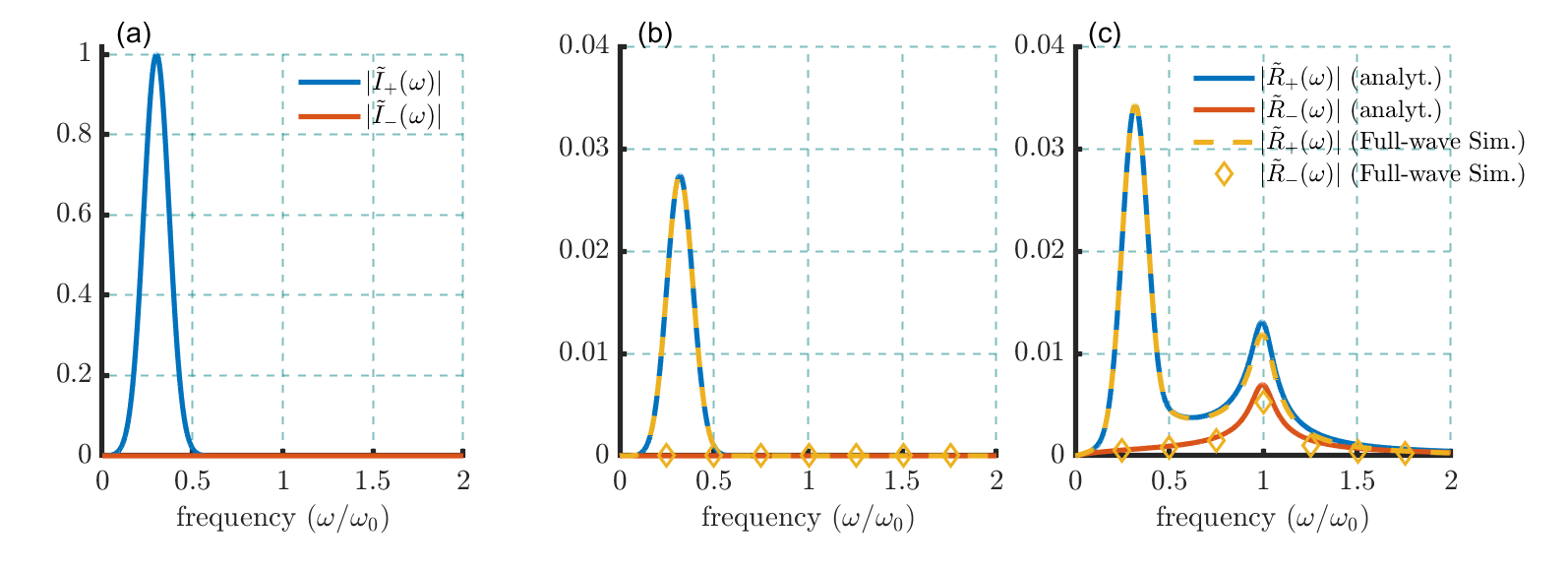}
 	\caption{Configuration and parameters as in Figure \ref{MT-fig2}. Comparison between analytic and full-wave results.}
 	\label{fig1S}
 \end{figure*}

\section{Semi-analytic approach for arbitrary modulation of the plasma frequency}

As discussed in the main text, electromagnetic scattering at the space-time interface can be treated analytically in cases where the plasma frequency undergoes an abrupt temporal transition. In more general scenarios, where the plasma frequency varies arbitrarily over time, the dynamics of the surface polarization $\bm{\mathcal{P}}(t)$ are still governed by Equation (\ref{MT-Lor_2}), which does not admit an analytic solution. 
In these scenarios, we resort to a semi-analytic approach, based on the numerical solution of Equation (\ref{MT-Lor_2}). To this aim, we employ MATLAB’s built-in ordinary differential equation solver \texttt{ode45} (MATLAB).
Once this numerical solution is obtained, the reflected and transmitted electric fields can be computed using Equations (\ref{MT-TR}).

\begin{figure*}
 	\includegraphics[width=1\textwidth]{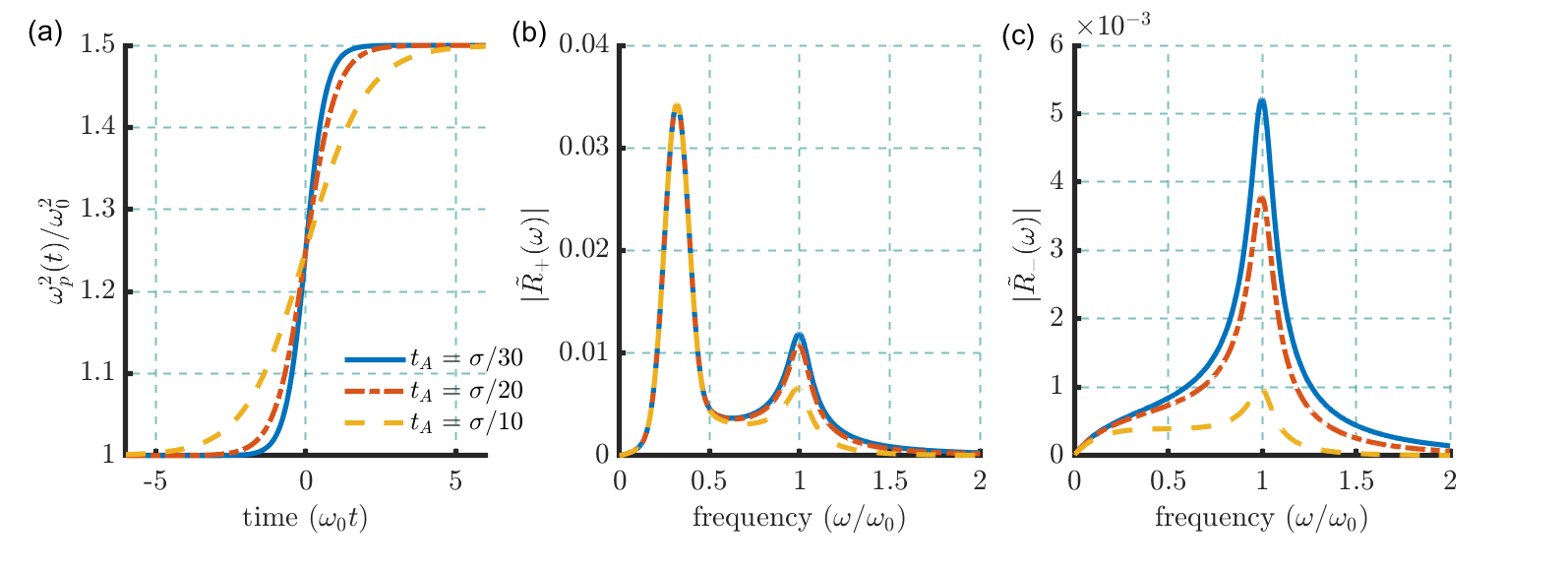}
 	\caption{Configuration and parameters as in Figure  \ref{MT-fig2}, but assuming a soft switching as described in Equation (\ref{SS}). a) Temporal profiles of $\omega_p^2(t)/\omega_0^2$, for different values of $t_A$. b,c) Corresponding positive- and negative-spin spectra  ($\tilde{R}_{+}$ and $\tilde{R}_{-}$, respectively) for the reflected (backward) wave. All spectra are normalized with respect to the peak value of the incident one.}
 	\label{fig2S}
 \end{figure*}

 This approach has been used to generate the results presented in Figures \ref{MT-fig4}, \ref{MT-fig5}, \ref{fig2S}, and \ref{fig3S}.

%
%

\section{Additional results}
 \begin{figure*}
 	\includegraphics[width=1\textwidth]{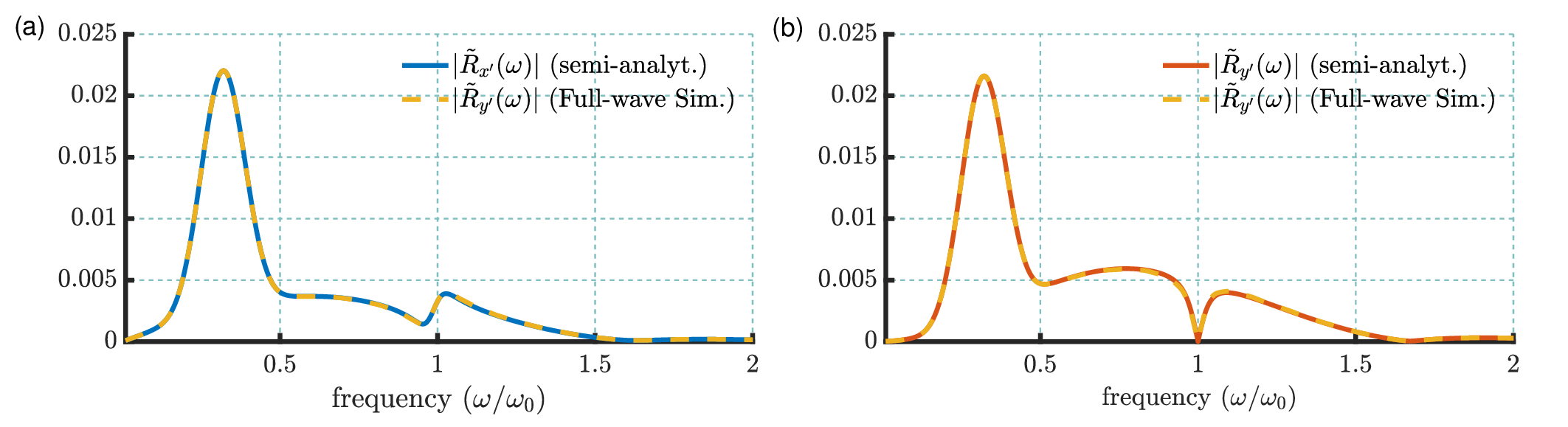}
 	\caption{Configuration and parameters as in Figure \ref{MT-fig5}b. Comparison between semi-analytic and full-wave results}
 	\label{fig3S}
 \end{figure*}

Figure \ref{fig1S} presents a comparison between the reflection spectra $\tilde{R}_{\pm}(\omega)$ obtained analytically in Figure \ref{MT-fig2} and those computed via full-wave simulations. Figures \ref{fig1S}b and \ref{fig1S}c correspond to the cases $A_2 = A_1$ and $A_2 = 1.5 A_1$ (with $A_1 = 1$), respectively. In the full-wave simulations, the plasma frequency undergoes a smooth transition described by
\begin{equation}
    A(t) = 0.5(A_1 + A_2) + 0.5(A_2 - A_1) \tanh\left(\frac{t}{t_A}\right),
    \label{SS}
\end{equation}
with a characteristic switching time $t_A = \sigma / 30$. As shown in Figure \ref{fig1S}, the numerical results obtained with this soft switching profile are in excellent agreement with the analytic predictions, validating the robustness of the model.

Figure \ref{fig2S} shows the reflection spectra for different values of the characteristic switching time $t_A$. As $t_A$ increases, the efficiency of frequency generation diminishes, leading to a corresponding decrease in polarization conversion.

For completeness, Figure \ref{fig3S} compares the reflection spectra presented in Figure \ref{MT-fig5} with those obtained from full-wave (COMSOL Multiphysics) simulations. The simulated results show excellent agreement with the semi-analytic predictions discussed in the main text, further confirming the occurrence of polarization conversion at the resonance frequency.

\medskip

%



